\documentclass[pre,aps,floatfix,superscriptaddress,twocolumn]{revtex4-2}
\usepackage{graphicx}
\usepackage{dcolumn}
\usepackage{latexsym}
\usepackage{hyperref}
\usepackage{amsmath, amsthm, amssymb}
\usepackage{epsfig}
\usepackage{bm}
\usepackage{geometry}
\usepackage[dvipsnames]{xcolor}
\usepackage{placeins}
\begin{document}


\title{Dynamics of phagocytosis through interplay of forces}

\author{Partha Sarathi Mondal}
\email[]{parthasarathimondal.rs.phy21@itbhu.ac.in}
\affiliation{Indian Institute of Technology (BHU) Varanasi, India 221005}

\author{Pawan Kumar Mishra}
\email[]{pawankumarmishra.rs.phy19@itbhu.ac.in}
\affiliation{Indian Institute of Technology (BHU) Varanasi, India 221005}

\author{Mitali Thorat}
\email[]{mitali.nthorat@gmail.com}
\affiliation{S.N. Bose National Centre for Basic Sciences, Kolkata, India 700098}

\author{Ananya Verma}
\email[]{averma14@syr.edu}
\affiliation{Department of Physics, Syracuse University, New York, United States, 13244}

\author{Shradha Mishra}
\email[]{smishra.phy@itbhu.ac.in}
\affiliation{Indian Institute of Technology (BHU) Varanasi, India 221005}


\date{\today}

\begin{abstract}
Phagocytosis is the process by which cells, which are $5–10$ times larger than the particle size, engulf particles measuring $\geq 0.5 \mu m$., holding substantial importance in various biological contexts ranging from the nutrient uptake of unicellular organisms to immune system of humans, animals etc. While the previous studies focused primarily on the mechanism of phagocytosis, in this study we have a taken a different route by studying the dynamics of the phagocytes in a system consisting of many bacteria and a small number of phagocytes. We put forward a minimalist framework that models bacteria and phagocytes as active and passive circular disks, respectively. The interactions are governed by directional forces: phagocytes are attracted toward bacteria, while bacteria experience a repulsive force in proximity to phagocytes. Bacteria are capable of reproduction at a fixed rate, and the balance between bacterial reproduction and phagocytic engulfment is governed by the interplay of the two opposing forces. In attraction-dominated regimes, bacterial populations decrease rapidly, while in repulsion-dominated regimes, bacterial clusters grow and impede phagocytes, often resulting in phagocyte trapping. Conversely, in attraction-dominated scenarios, only a few bacteria remain at later times, rendering the motion of the phagocytes diffusive. Further, the transition between the two regimes occurs through a regime of bi-stability. Our study further describes the dynamics of both species using the tools of  statistical analysis, offering insights into the internal dynamics of this system.
\end{abstract}

\maketitle

\begin{figure*} [hbt]
\centering
\includegraphics[width=0.98\textwidth]{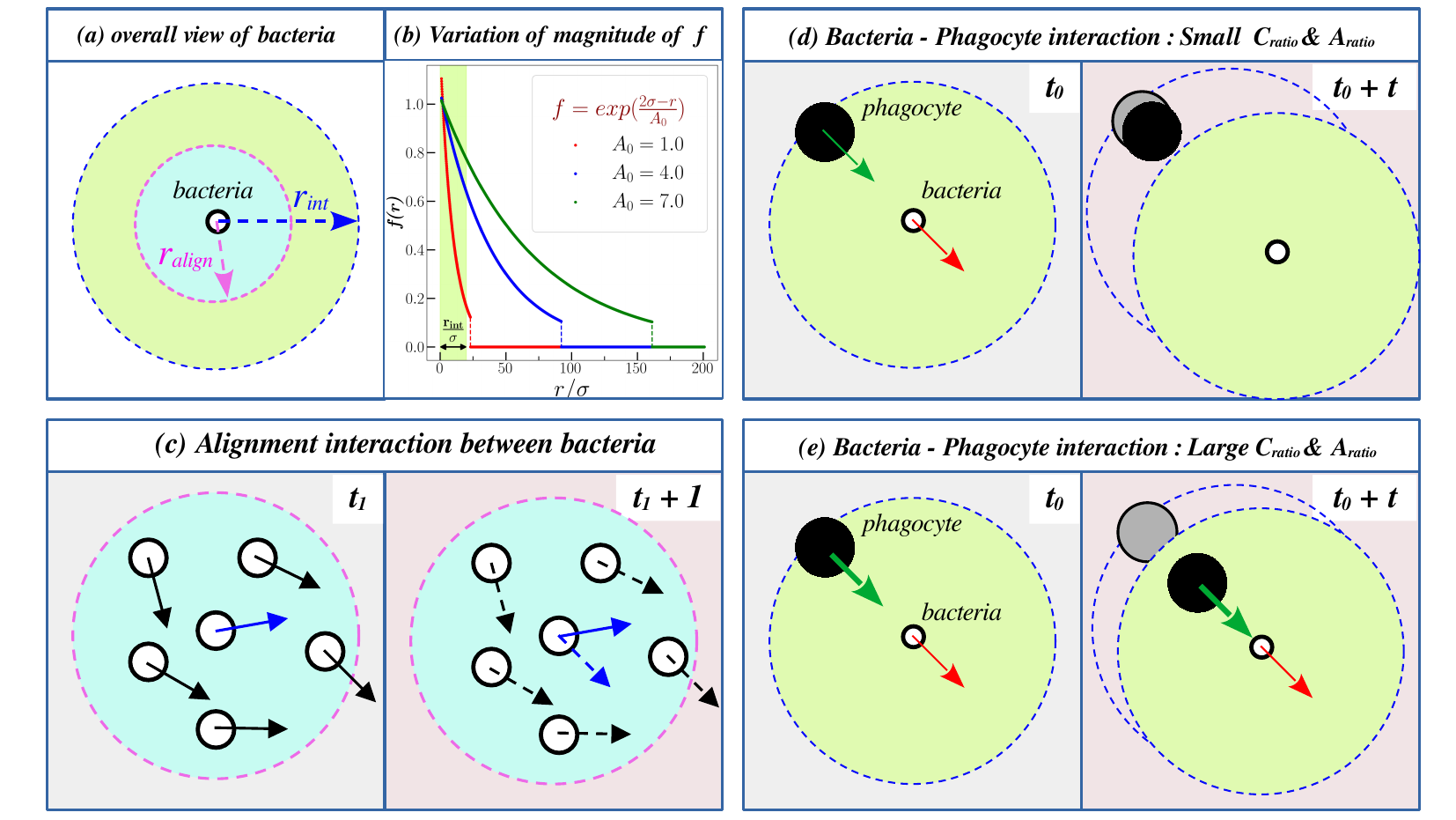}
\caption{(color online) The figure represents visual description of the main features of the model. Subplots (a) provides overall description of bacteria and it's properties. The subplot (b) shows the variation of force $f(r)$ for different values of $A_0$. Subplot (c) presents a cartoon of the alignment interaction between bacteria at two subsequent time steps. The self propulsion direction of $i^{th}$ bacteria is shown with blue color arrow and the the self-propulsion direction of other bacteria inside the interaction radius is shown with black arrow. The instantaneous self propulsion direction of a bacteria is shown with solid arrow and that at the previous time step is shown with dashed arrow. The subplots (d) and (e) showcases a cartoon of the interaction between bacteria and phagocyte at two times. The green arrow on phagocyte shows it's tendency to move towards the bacteria and the red arrow on bacteria shows it's tendency to move away from the phagocyte. The thickness of the arrows represents the relative strength of the forces in respective cases. The subplot (d) shows that for small $C_{ratio}$ and $A_{ratio}$ the strength of the two forces are comparable and the bacteria has larger probability to escape engulfment. The subplot (e) shows that for larger large $C_{ratio}$ and $A_{ratio}$ the attraction force on phagocyte is much stronger compared to the repulsion force on bacteria, and hence, the bacteria has much less probability to escape engulfment. In both the subplots (d) $\&$ (e), the 'grey' circles mark the position of phagocyte at time $t_0$ compared to that at time  $t_0 + t$.the frame at subsequent times is shifted in accordance to center of mass of bacteria.}
\label{fig:1}
\end{figure*}

\section{Introduction\label{secI}}

Active matter systems \cite{vicsek1995novel,bechinger2016active,bar2020self,vicsek2012collective} consist of agents which are self propelled by energy input at single particle level, and hence driving the system far from equilibrium. Consequently, these systems showcase characteristics which are not possible in their equilibrium counter parts \cite{toner2005hydrodynamics,marchetti2013hydrodynamics,ramaswamy2017active,das2020introduction}. Such systems have also been studied with birth and death \cite{toner2012birth,mishra2022active,jena2023ordering,jena2024spatio}. The complexity of the inherent dynamics of the system leaves its trace in the response of the system to some external perturbations like presence of foreign agents \cite{mondal2024dynamical,Mishra_2024,sampat2021polar}. Depending on the nature of these foreign agents the system can be categorized in different sub-classes. If the foreign agents are quenched in space and time, then they act as disorders to the agents constituting the active systems. The disordered systems \cite{mccandlish2012spontaneous,singh2021bond,chepizhko2013optimal,das2018polar} have received significant attention owing to their ability to mimic the complexities encountered by self propelled agents in natural environments. However, if the foreign agents are mobile, then the system is referred to as active passive mixture \cite{dolai2018phase}. Here, the interaction between the two species play an important role in determining the behavior of the system. Moreover, if the foreign agents are motile, then the interplay of the inter- and intra-species interactions as well as the self propulsion of the two species adds another layer of complexity in the system's behavior.\\
Many natural systems can be molded in to the structure of these active-passive and active-active mixtures. These systems are prevalent across various length scales. On the microscopic scale one popular example of a process which involves the dynamics of the two different species of motile agents is Phagocytosis \cite{uribe2020phagocytosis}. From biological point of view this process hold significant importance from nutrient uptake of unicellular organisms to immune systems of many multicellular organisms including humans, animals etc. Consequently, it has received significant attention from experimental point of view in order to reveal different controlling parameters of the process and it's characteristics features. However, the limitedness of the accessible values of the control parameters sets restrictions on the in vitro experimental studies. This emphasizes the importance of numerical studies. However, previous numerical \cite{richards2014mechanism,herant2006mechanics,sadhu2023theoretical} studies were focused on the examining the different stages of the engulfment process of an external substance (e.g. microorganisms, food particles etc.) by a phagocyte. Using Phase Field Modeling technique \cite{hamed2021three}, the previous studies have been successful in describing the physics of the different stages of the process \cite{winkler2024physical}. However, in a multi agent system the timescale of a single engulfment process is much smaller compared to other relevant timescales. Therefore, it is important to explore the possible behavior of a multi agent system of phagocytes and external substances. To our knowledge, there are no such numerical studies addressing this. Although the actual process involves numerous biological factors, it is crucial to avoid overburdening numerical models with excessive details and parameters, as this can obscure a clear understanding of the system’s behavior across its parameter space. Minimal models have proven effective in capturing the dynamics of complex systems by relying on carefully selected interactions and control parameters that encapsulate the system’s essential complexities.\\ 
In this study we present a minimal model to study phagocytosis in a multi agent system. The external substances are referred to as bacteria and are modeled as active Brownian particles, whereas the phagocytes are modeled as passive Brownian particles. The interspecies interaction between bacteria and phagocytes is governed by two forces: a repulsive force that drives bacteria away from phagocytes and an attractive force that draws phagocytes toward bacteria. Additionally, bacteria can reproduce with a certain probability, adding a dynamic component to the system. The relative range and the relative strength of the two forces  serves as the control parameters of the model.\\
The main findings of this study are as follows: (i) We observe a competition between bacterial reproduction and phagocytic engulfment, with the system reaching one of two distinct outcomes based on the dominant process for a given set of parameters. In one scenario, phagocytosis is ineffective, allowing bacteria to persist and multiply. In the other, phagocytes successfully engulf nearly all bacteria. (ii) Near the phase boundary between these states, the system exhibits bistability, where both outcomes are possible.\\
The organization of the rest of the manuscript is as follows: the Section \ref{sec:mod} describes our model and provides details about our parameter space. Then our results are discussed in Section \ref{sec:res}. Finally, we close with a summery of the study and the future scopes of the model in Section \ref{sec:dis}. \\


\renewcommand{\arraystretch}{1.5} 
\setlength{\tabcolsep}{10pt} 

\begin{table*}
    \caption{The table presents the parameters of the model. The length and time scales of the model are rescaled with respect to $\sigma_b$ and $\tau$.}
    \centering
    \begin{tabular}{|c|c|c|c|}
    \hline
    \textbf{Parameter} & \textbf{Symbol} & \textbf{Value} & \textbf{Rescaled value}\\
    \hline
    Number of bacteria at $t = 0$ & $n_{b,0}$ & 500 & -\\
    \hline
    Number of phagocytes at $t = 0$ & $N_p$ & 7 & -\\
    \hline
    Radius of bacteria & $\sigma$ &  0.10 & -\\
    \hline
    Radius of phagocyte & $\sigma_p$ & 0.60 & 6$\sigma_b$\\
    \hline
    self propulsion speed of bacteria & $v_0$ &  0.1 & -\\
    \hline
    Interaction range of bacteria & $R_b$ & 1.0 & 10$\sigma_b$\\
    \hline
    Interaction range of phagocyte & $R_{bo}$ & 1.0 & 10$\sigma_b$\\
    \hline
    Thickness of receptor layer on phagocyte surface & $\Delta R_{rec}$ & 0.20 & 2$\sigma_b$\\
    \hline
    Strength of hardcore interaction force & $C_{core}$ & 30.0 & -\\
    \hline
    strength of noise & $\eta$ & 0.01 & -\\
    \hline    
    time step for simulation & $\delta t$ & 0.01 & 0.01$\tau$\\
    \hline
    Diffusivity of Phagocyte & $D$ & 0.001 & -\\
    \hline
    \end{tabular}
    \label{tab:tab1}
\end{table*}

\section{Model}\label{sec:mod}
{\textit{ Dynamical equations for bacteria}} : 
Bacteria are modelled as self propelled disks with radius $\sigma$ and self propulsion velocity $v_{0}$. Bacteria are characterized by position  and velocity of the center of the disk denoted by $\bold{r}_{b}(t)$ and $\bold{V}_{b}(t)$, respectively. At each time step bacteria can reproduce with a probability $P_{rep}$.\\
Instantaneous velocity of bacteria at a  given time has two contributions: bacteria, due to their self propelled nature, try to move along the direction of it's head and align with the other particles in it's neighbourhood. This inherent alignment tendency of bacteria is disturbed by their interaction of other forms with particles in their neighbourhood. The instantaneous velocity of the $i^{th}$ bacteria at time $(t+1)$ is given by,
\begin{equation}
    \boldsymbol{V}_{i,b}(t+1)=\boldsymbol{V}_{i,b}^{align}(t+1)+\boldsymbol{V}_{i,b}^{int}(t+1) 
    \label{equn:1}
\end{equation}
where, the $1^{st}$ term represents noisy alignment of each bacteria with other bacteria it's alignment range denoted by $R_b$. The $2^{nd}$ term accounts for the contribution to velocity of the $i^{th}$ bacteria coming from it's interactions of with other bacteria and phagocytes inside it's interaction range denoted by $r_{int}$.\\
The alignment term is calculated as: 
\begin{center}
$\boldsymbol{V}_{i,b}^{align}(t+\Delta t) = v_{0}(1-\lambda_1 \rho_1)\hat{e}_{i,b}(t+\Delta t)$
\end{center}
\begin{center}
$\hat{e}_{i,b}(t+\Delta t)\equiv \bigg(cos[\theta_{i,b}(t+\Delta t)],sin[\theta_{i,b}(t+\Delta t)] \bigg)$\\
$\theta_{i,b}(t+\Delta t)=\bigg<\theta(t) \bigg>_{R_{b}}+\eta_{i,b}(t)$
\end{center}
where, $v_{0}$ is the self propulsion speed of bacteria, and $\eta_{i,b}(t)$ represents a delta correlated white noise with zero mean chosen uniformly form the range $[\frac{\eta}{2},\frac{\eta}{2}]$. The term $(1-\lambda_1 \rho_1)$ sets a feedback loop between the instantaneous velocity of bacteria and local density of bacteria inside it's interaction radius, $\rho_1$. This density dependence of bacteria promotes the clustering of bacteria.\\
The interaction term is defined as,
\begin{center}
 $\boldsymbol{V}_{i,b}^{int}(t+1) = \boldsymbol{V}_{i,bb}^{rep}(t+1) + \boldsymbol{V}_{i,bp}^{rep}(t+1) $   
\end{center}
where, the two terms on the right hand side accounts for the interaction of a bacteria with other bacteria and phagocytes inside it's interaction radius, respectively. The $1^{st}$ term on right hand side expresses the volume exclusion interaction between bacteria and is given by,
\begin{equation}
    \boldsymbol{V}_{i,bb}^{rep}= C_{core} \sum_{|\boldsymbol{r}_{ij}| \in \Omega_i} g(\sigma_i+\sigma_j - |\boldsymbol{r}_{ij}|) \hat{e}_{ij}
    \label{equn:2}
\end{equation}
where, $g(x)=\vert x \vert$ if $x>0$ and 0 if $x<0$. \\
The $2^{nd}$ term on right hand side is a repulsive force that models the tendency to move away from the phagocytes inside it's interaction range to escape the engulfment. The contribution of this repulsive force to velocity is given by,
\begin{equation}
    \boldsymbol{V}_{i,bp}^{rep}= C_{rep} \sum_{|\boldsymbol{r}_{ij}| \in \Omega_i} f(|\boldsymbol{r}_{ij}|)\hat{e}
    \label{equn:3}
\end{equation}
where, $\Omega_i$ denote the subset of particles that are inside the interaction range of the $i^{th}$ bacteria. The strength of the repulsion is denoted by $C_{rep}$ ,and the form of the interaction is given by,
\begin{equation}
    f(r,A_0) = exp \bigg[\frac{\sigma_i+\sigma_j - |\boldsymbol{r}_{ij}|}{A_0}\bigg]
    \label{equn:4}
\end{equation}
where, the parameter $A_0$ decides how fast $f(r)$ decays with distance. The variation of $f(r)$ with distance is shown in FIG.\ref{fig:1}(b). In Eq.(\ref{equn:2}-\ref{equn:4}), the subscripts $i$, $j$ denotes two interacting particles and  $\boldsymbol{r}_{ij} = \boldsymbol{r}_i - \boldsymbol{r}_j$. $\hat{e} = \frac{\boldsymbol{r}_{ij}}{|\boldsymbol{r}_{ij}|}$ is the unit vector along the line from the center of the $j^{th}$ particle to $i^{th}$ particle.

Position update equation of $i^{th}$ bacteria is given by,
\begin{equation}
    \boldsymbol{r}_{i,b}(t+1)=\boldsymbol{r}_{i,b}(t)+\Delta t \boldsymbol{V}_{i,b}(t+1)
    \label{equn:5}
\end{equation}
where, $\Delta t$ is the time step of simulation.\\
In natural systems, bacteria reproduce through cell division. In our model, this process is implemented in an event-driven manner as described below: 
\begin{enumerate}
    \item[(i)] At each time step, a uniform random number $r \in [0,1]$ is generated for each bacterium $i$. The $i^{th}$ bacterium can reproduce if $r > P_{rep}$, where $P_{rep}$ is the reproduction probability.
    \item[(ii)] The position $(x,y)$ of the new bacteria is chosen as $\boldsymbol{r}_{new} = \boldsymbol{r}_{i,b} + 2\sigma (\cos\phi ,\sin\phi)$.
    where, $\phi \in [0,2\pi]$ is chosen uniformly. The new position is assigned to the offspring only if it does not overlap with any existing bacterium.
\end{enumerate}
In our numerical simulation we set the reproduction probability, $P_{rep} = 0.99$. Additionally, bacteria do not undergo natural death; they are removed from the system only if engulfed by a phagocyte.\\

\begin{figure*} [hbt]
\centering
\includegraphics[width=0.98\textwidth]{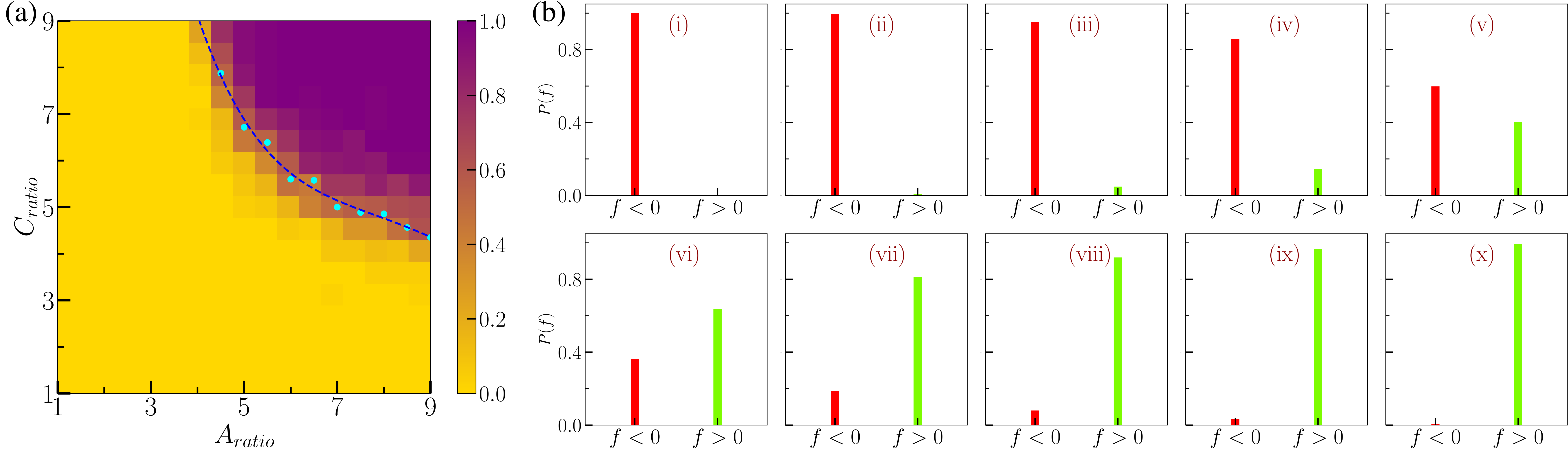} 
\caption{(color online) The figure depicts the behavior of the system for different $C_{ratio}$ values across the phase boundary. (a) The variation of the number of bacteria in the system, $n_b(t)$ vs. $t$ for different $C_{ratio}$ values across the boundary between $f>0$ and $f<0$ phases for $A_{ratio} = 6.0$; (b) The statistical behavior of the system over different independent realizations for different $C_{ratio}$ values. For $C_{ratio}$ values deep in $f>0$ or $f<0$ region we observe that behavior of the behavior of the system in different ensembles is consistent, whereas for $C_{ratio}$ values close to the phase boundary the system shows bi-stability i.e. the system converges to $f>0$ state for some ensembles and to $f<0$ state for some other ensembles for a given $C_{ratio}$ value. The results shown are obtained for $A_{ratio} = 6$. The results obtained for $A_{ratio} = 7$ are qualitatively similar. The rest of the parameters are same as listed in Table.\ref{tab:tab1}.}
\label{fig:2}
\end{figure*}


{\em Dynamical equations for phagocytes} : 
Phagocytes are modelled as passive disks of radius $\sigma_p$ and the have a receptor layer of thickness $r_{rec}$ on their surface. Phagocytes are characterized by the position and velocity of the center of the disks denoted by $\bold{r}_{p}(t)$ and $\bold{V}_{p}(t)$, respectively. For a phagocyte, the velocity vector represents the instantaneous direction of motion of the phagocyte, which is decided by it's interaction with the other particles inside the interaction range.\\
The instantaneous velocity of the $i^{th}$ phagocyte at time $(t+1)$ is given by,
\begin{equation}
    \boldsymbol{V}_{i,p}(t+1)=\boldsymbol{V}_{i,pb}^{att}(t+1)+\boldsymbol{V}_{i,pp}^{ex}(t+1)
    \label{equn:6}
\end{equation}
where, the two terms on the right hand side accounts for phagocyte-bacteria and phagocyte-phagocyte interactions, respectively. The phagocyte-phagocyte interaction adheres to volume exclusion, denoted by the $2^{nd}$ term on the right hand side, and the form of the interaction is similar to that given by Eq.\ref{equn:2}. The phagocyte-bacteria interaction is given by a exponentially decaying attraction force which mimics the chemotatic motion of phagocytes following the gradient of the chemical secreted by bacteria. The contribution of this attractive interaction to the velocity of $i^{th}$ phagocyte is given by,
\begin{equation}
    \boldsymbol{V}_{i,pb}^{att} = C'_{att} \sum_{j} f_{1}(|\boldsymbol{r}_{ij}|)\hat{e}_{1}
    \label{equn:7}
\end{equation}
where, $\hat{e}_{1}$ is the unit vector along the line from the center of the $i^{th}$ phagocyte to $j^{th}$ bacteria, and $C'_{att} = C_{att}(1-\lambda \rho_2)$ is the density dependent attraction strength. $f_1(r)$ decides the nature of variation of the attractive force such that,
\begin{equation}
f_1(r) =
\begin{cases} 
    f(r,A_1), & \text{if } f(r,A_1) > \frac{1}{10} f(0,A_1) \\
    0, & \text{otherwise}
\end{cases}
\label{equn:8}
\end{equation}

Further, when a bacteria is inside the range of the receptor layer of a phagocyte (i.e. $|{\bf r}_{ij}| < (\sigma_p + r_{rec})$), it gets engulfed. In our model, this is implemented by adding an extra term in the right hand side of Eq.\ref{equn:7}, which has the following form,
\begin{equation*}
   V_{pb}^{en} = C'_{core} g(\sigma_p+r_{rec}+\sigma_ - |\boldsymbol{r}_{ij}|) \hat{e}_{1}
\end{equation*}
where, $|\boldsymbol{r}_{ij}|$ is the distance between the $i^{th}$ phagocyte and $j^{th}$ bacteria. 

The position update equation of $i^{th}$ phagocyte is given by,
\begin{equation}
    \boldsymbol{r}_{i,p}(t+1)=\boldsymbol{r}_{i,p}(t)+\Delta t.\boldsymbol{V}_{i,p}(t+1)+ D \sqrt{dt} \boldsymbol{\zeta}_i
    \label{equn:9}
\end{equation}
where, the $3^{rd}$ term describes the diffusion in the motion of phagocytes and $D$ is the coefficient of diffusion, $\zeta_i \equiv (\cos \phi_i,\sin \phi_i)$ is a random unit vector with $\phi_i \in [0,2\pi]$.

\begin{figure*}[hbt]
    \centering
    \includegraphics[width=0.999\linewidth]{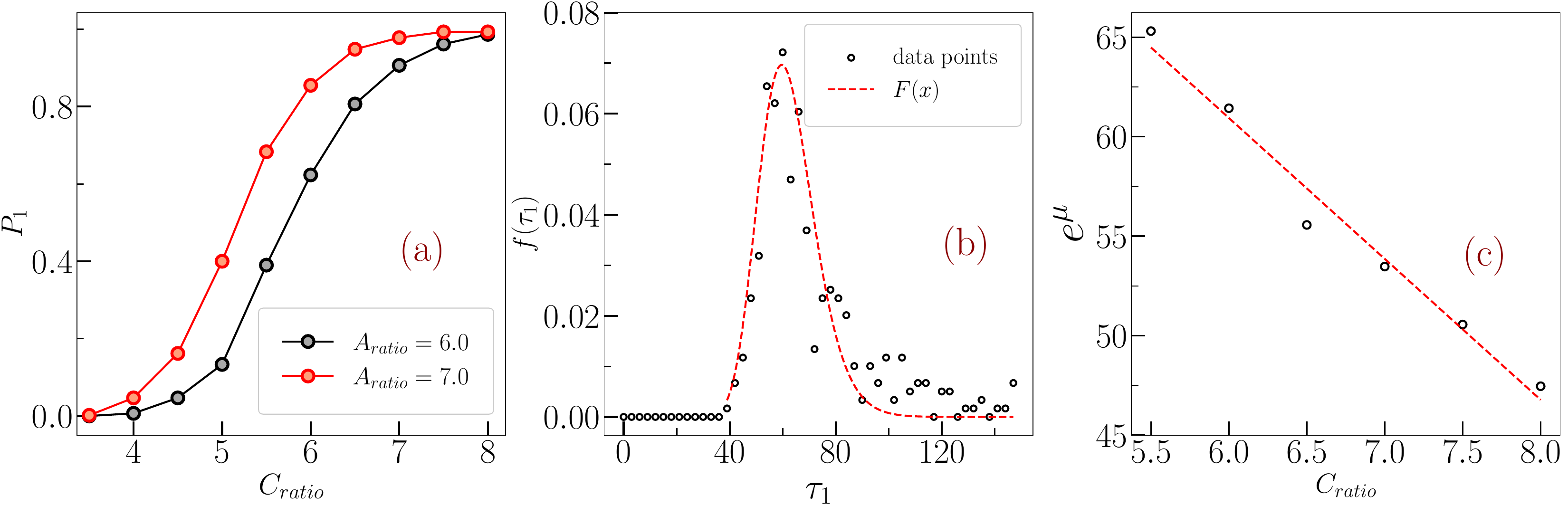}
    \caption{(color online) The figure depicts: (a) the variation of $P_1$ with $C_{ratio}$; (b) the PDF of $\tau_1$ for $C_{ratio} = 6.0$ shown with `\textbf{o}' symbol. The red dashed line shows the fit with log-normal distribution; (c) the  variation of the mean of the distribution, $e^{\mu}$, with $C_{ratio}$ values.}
    \label{fig:3}
\end{figure*}

\subsection*{Parameters and Simulation Details}
In this model, the steady state behavior is decided by the tussle between the bacteria-phagocyte repulsion and the phagocyte-bacteria attraction. To delve deeper into this, we define the control parameters as, $\lambda$, $C_{ratio} = \frac{C_{att}}{C_{rep}}$ and $A_{ratio} = \frac{A_1}{A_0}$, which controls the relative strength and the decay rate of the two forces, respectively. In our simulation we considered three different values of the parameter $\lambda$, $\lambda$ $=$ $0.20$, $0.40$, and $0.50$. The results shown below are obtained $\lambda$ $=$ $0.50$,. However, we ensure that the results for other values of $\lambda$ are qualitatively similar. Rest of the parameters are listed in Table.\ref{tab:tab1}.\\
The intrinsic length and time scale in the system are given by the radius of the bacteria $\sigma_b$ and $\tau = \frac{\sigma_b}{v_0} = 1.0$, respectively. The Equns.\ref{equn:5} and \ref{equn:9} are rescaled with this intrinsic length scale and time scale.\\
We begin with a very low density of bacteria and very small number of phagocytes. The system is simulated for a total of 15000 time steps, during which we calculate our various observable quantities. A single simulation step is counted when the state of all the particles (bacteria + phagocytes), as well as total number of particles in the system are updated. A total of 600-1000 independent realisations (different initial distribution of particles as well as different configuration of the noise) are considered for better statistics. Further, during one single simulation run, we stop the iteration process if the number of bacteria exceed 5 times its initial value  or there are no bacteria left in the system.\\

\section{Results}\label{sec:res}
The number of bacteria in the system is decided by the trade off between the two counteracting contributions: the reproduction of bacteria which increases the number and the engulfment of bacteria by phagocyte which decreases the number. The contribution from the former depends on the reproduction rate (kept fixed in our model), while the contribution of the latter depends on how efficiently the phagocytes can engulf bacteria. The dominant one of the two aforementioned mechanisms decides the fate of the system for a given set of parameters. To track which mechanism is dominant for a given set of parameters we calculate $f(t) = \frac{(n_{b,0} - n_b(t))}{n_{b,0}}$,  where, $n_b(t)$, and $n_{b,0}$ are the number of bacteria in the system at time t, and t=0, respectively. \vspace{0.1 cm}

{\em Phase Diagram} : First, we present a complete phase diagram of the system in FIG.\ref{fig:2}(a). The phase diagram depicts the value of the probability of $f>0$, denoted by $P_{+}$, for each set of parameters. The details of the calculation of the probability, $P_{+}$, is described below.\\
For each set of parameters we consider $800$ independent runs. Each run consists of $25 \times 10^3$ time steps. However, a run is terminated if the number of bacteria in the system increases beyond $5$ times the initial number or it decreases to zero. For each run we track the value of $f$ over the entire run and use the value of $f$ averaged over last 200 time steps to determine $f > 0$ or $f < 0$ for the run. The probability $P_{+}$ is defined for the given set of parameters is defined as, $P_{+} = \frac{N_{f>0}}{N_{tot}}$, where, $N_{f>0}$ denotes the number of ensembles for which $f>0$ is obtained and $N_{tot}$ is the total number of ensembles. In the phase diagram shown in FIG.\ref{fig:2}(a), the colorbar represents the value of $P_{+}$ for the given set of parameters. For a given set of parameters, a value $P(f) \approx 0$ represents the ineffectiveness of the Phagocytosis wherein engulfment rate is dominated over by reproduction rate, whereas a value $P(f) \approx 1$ represents dominance of the engulfment rate over reproduction rate leading to efficient phagocytosis in the system. From the phase diagram it is evident that for relatively small values of $C_{ratio}$ and $A_{ratio}$, when the range and rate of decay of the repulsive and attractive force are comparable, the system exhibits $f<0$ state. However, when both $C_{ratio}$ and $A_{ratio}$ are large, the final state of the system exhibits $f>0$. However, the transition between the two phases is not sharp as indicated by the intermediate value of $P_{+}$ near the boundary between the $f>0$ and $f<0$ phases (marked by blue dashed line). 

Next we explore the characteristic behavior of the system in the vicinity the phase boundary. Below, we show results for $A_{ratio} = 6.0$ and $C_{ratio} \in [3.5, 8.0]$ wherein we move along a vertical line across the phase boundary. However, we have confirmed that on moving across any other vertical or horizontal line across the phase boundary one observes the similar qualitative changes in the behavior of the system as discussed below.

\begin{figure*}[hbt]
    \centering
    \includegraphics[width=0.999\linewidth]{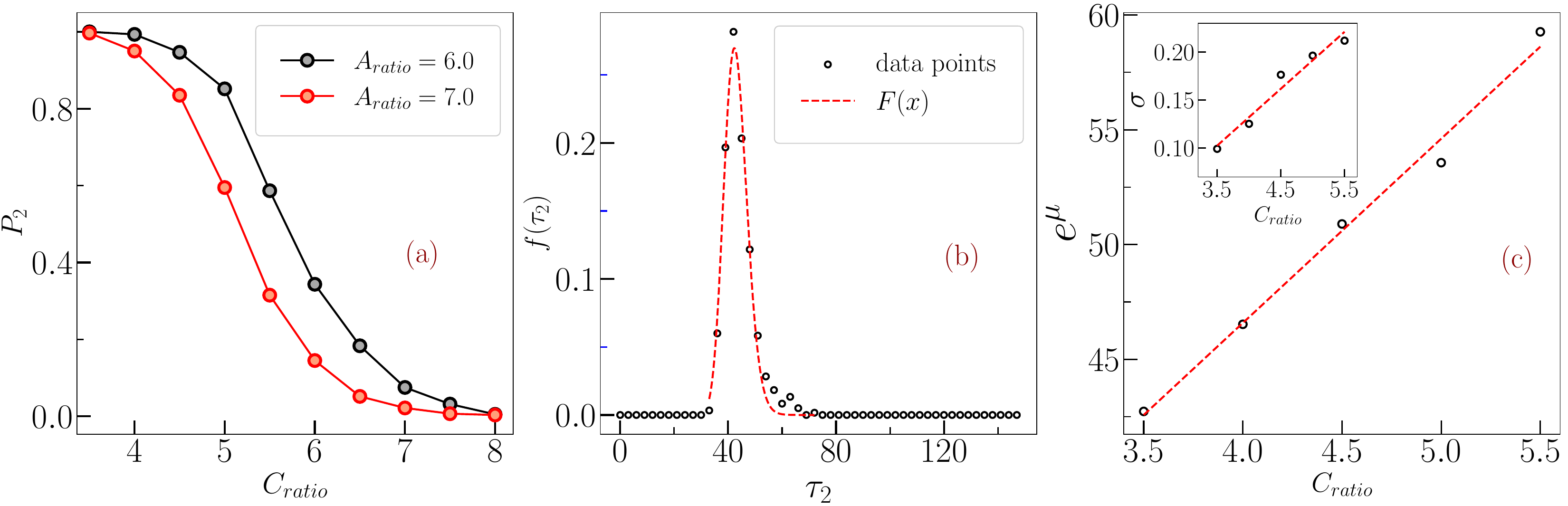}
    \caption{(color online) The figure depicts: (a) the variation of $P_2$ with $C_{ratio}$; (b) the PDF of $\tau_2$ for $C_{ratio} = 6.0$ shown with `\textbf{o}' symbol. The red dashed line shows the fit with log-normal distribution; (c) the variation of the mean of the distribution, $e^{\mu}$, with $C_{ratio}$ in the main plot. The inset shows the variation of the standard deviation, $\sigma$, with $C_{ratio}$.  }
    \label{fig:4}
\end{figure*}

{\em Bi-stability in the vicinity of the phase boundary :} As we approach the phase boundary from one side, the system enters a regime where $P_{+}$ takes intermediate values. To investigate the characteristics of this transition region, we perform a statistical analysis of the system's behavior across multiple ensembles. We calculate probability of $P_{+}$ and $P_{-}$ for $A_{ratio} = 6.0$ and different values of $C_{ratio} \in [3.5, 8.0]$ as shown in FIG.\ref{fig:2}(b). The probability is calculated over $800$ independent runs (ensembles), as described previously. It is evident that, far away from the phase boundary, the system settles to either $f>0$ or $f<0$ state depending on the parameter values with all the ensembles being equivalent for a given set of parameters. However, in the vicinity of phase boundary we observe that both $P_{+}$ and $P_{-}$ take finite nonzero value, in accordence with the normalization condition $P_{+} + P_{-} = 1$. At parameter values closest to the phase boundary, $P_{+}$ and $P_{-}$ are approximately equal, indicating an equal likelihood of $f>0$ and $f<0$ states states for this parameter range. Thus, near the phase boundary, the system exhibits bistability.\\
Further, to delve even deeper into the behavior of the system across the phase boundary we calculate the probability of the number of bacteria in the system to reach (i) $5\%$ or less of the initial number, defined as $P_1 = \frac{N_1}{N_{tot}}$, and (ii) $5$ times or more of the initial number, defined as $P_2 = \frac{N_2}{N_{tot}}$ , where $N_1$ and $N_2$ are the number of ensembles for which the conditions $n_b(t) \le 5\% n_{b,0}$ and ${n_b(t) \ge 5n_{b,0}}$, respectively, are satisfied. The details of calculation of the probabilities, $P_1$ $\&$ $P_2$ are same as described above.

The behavior of $P_1$ on varying $C_{ratio}$ is shown in FIG.\ref{fig:3}(a) for two values of $A_{ratio}$. On increasing $C_{ratio}$, $P_1$ gradually increases. This behavior is consistent for different values of $A_{ratio}$. Further, for each ensemble satisfying the condition $n_b(t) \le 5\% n_{b,0}$, we calculate the time, denoted by $\tau_1$, when the cut-off is reached for the first time . We calculate the probability distribution function (PDF) of $\tau_1$, denoted by $f(\tau_1)$ and,  fit it to a log-normal distribution given by:
\begin{equation}
    F(x) = \frac{1}{\sigma x \sqrt{2\pi}} exp\bigg(-\frac{(lnx - \mu)^2}{2\sigma^2}\bigg)
    \label{eq:6}
\end{equation}
where, $e^{\mu}$  and $\sigma$ denoted the mean and standard deviation of the variable $x$. In FIG.\ref{fig:3}(b) we show the PDF of $\tau_1$ and the fit with log normal distribution for $C_{ratio} = 6.0$. The mean of the distribution, $e^{\mu}$, decreases with increase in $C_{ratio}$, as shown in FIG.\ref{fig:3}(c), whereas the standard deviation of the distribution, $\sigma$, remains the same for all the $C_{ratio}$ values considered. These results show that as we move deep in to the $f > 0$ regime, the number of bacteria decays faster with time.\\
The variation of $P_2$ with $C_{ratio}$ is shown in FIG.\ref{fig:4}(a) for two $A_{ratio}$ values. With increase of $C_{ratio}$, $P_2$ decreases. With increase of $A_{ratio}$, $P_2$ decays faster. Further, for each ensemble satisfying the condition $n_b(t) \ge 5n_{b,0}$, we calculated the time, denoted by $\tau_2$, when the cut-off is reached for the first time . We then compute the probability distribution function (PDF) of $\tau_2$, denoted by $f(\tau_2)$ and fit it with log-normal distribution given by Eq.\ref{eq:6}. The fit with log-normal distribution is shown in FIG.\ref{fig:4}(b) for $C_{ratio}=3.5$. The mean, $e^{\mu}$ and standard deviation, $\sigma$, of $f(\tau_2)$ obtained from the fit is plotted in FIG.\ref{fig:4}(b). Both mean and standard deviation increases with increase of $C_{ratio}$.\\

These observations confirm that the transition from the $f < 0$ to the $f > 0$ region gradual and the system passes through a state of bi-stability wherein the both the states are equally possible. \\
Next we delve deeper in to the difference in the  static and dynamical characteristics of the system in the $f > 0$ and $f < 0$ regions in the phase diagram. In the following, we restrict our analysis to $C_{ratio}$ values away from the phase boundary. Since, in the close vicinity of the phase boundary different independent runs for a same set of control parameters are not equivalent as a consequence of bi-stability. The results for $f > 0$ and $f < 0$ phases shown below correspond to $C_{ratio}$ values $[3.5, 4.0, 4.5]$, and $[8.0, 8.5, 9.0]$, respectively with $A_{ratio} = 6.0$. However, we have confirmed that the results for any other set of parameter values deep in $f > 0$ or $f < 0$ phase are qualitatively same as the following analysis in the respective cases.\\

\begin{figure*}[hbt]
    \centering
    \includegraphics[width=0.995\linewidth]{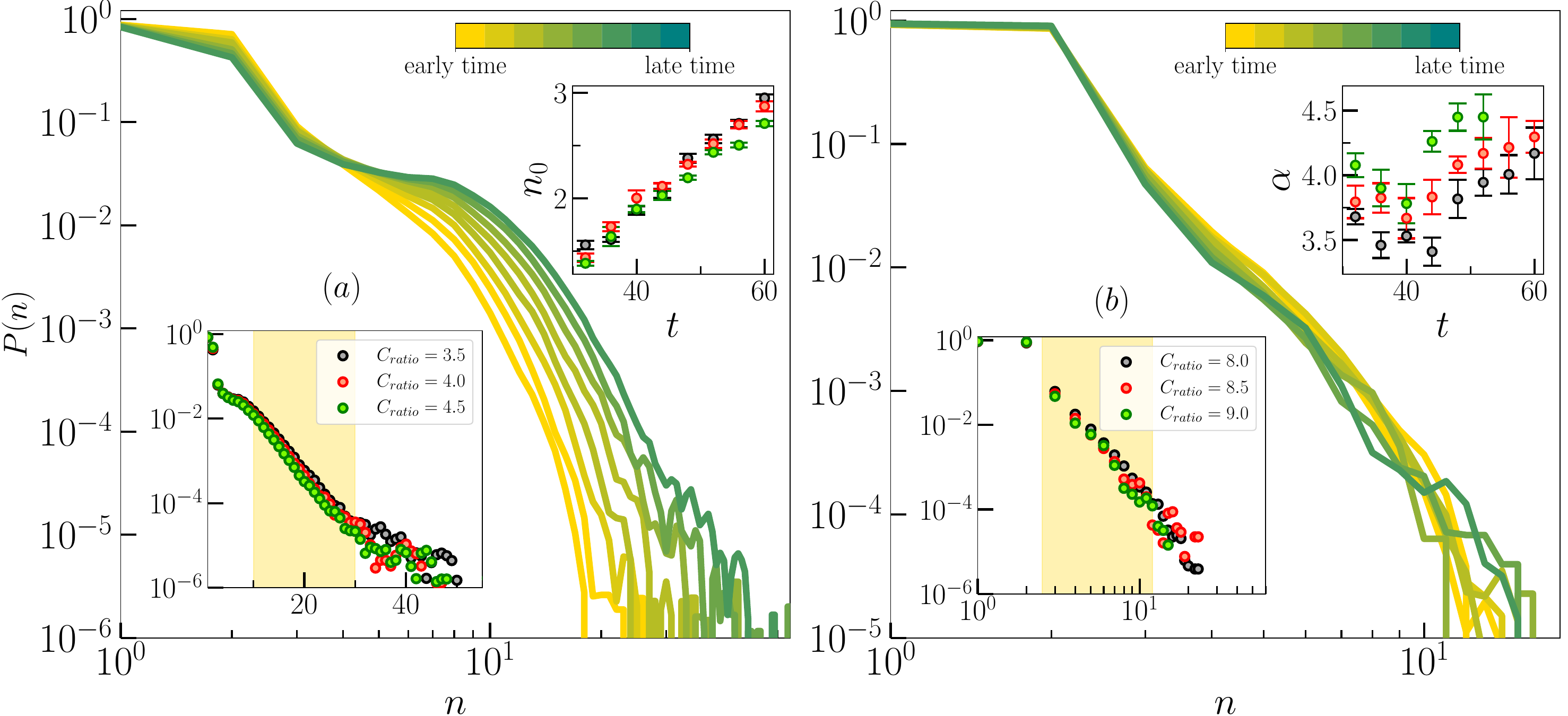}
    \caption{(color online) The figure displays the behavior of cluster size distribution (CSD), $p(n)$ $vs.$ $n$, for $C_{ratio}$ values in $f < 0$, panel (a), and $f > 0$, panel (b), region. In panel (a), the main frame shows the plot of $p(n)$ $vs.$ $n$ in log-log scale at different times indicated by the color bar for $C_{ratio}=3.5$. The bottom-left inset shows the plot of $p(n)$ $vs.$ $n$ in log-y scale at late time for different values of $C_{ratio}$. The highlighted region in the inset shows the exponential decay of $P(n)$ for intermediate values of $n$ : $P(n) \sim exp(-n/n_0)$. The top-right inset shows the variation of $n_0$ with time for different values of $C_{ratio}$. The color code is same for both the insets. In panel (b), the main frame shows the plot of $p(n)$ $vs.$ $n$ in log-log scale at different times indicated by the color bar for $C_{ratio}=8.5$. The bottom-left inset shows the plot of $p(n)$ $vs.$ $n$ in log-log scale at late time for different values of $C_{ratio}$. The highlighted region in the inset shows the powerlaw decay of $P(n)$ for intermediate values of $n$. The top-right inset shows the variation of $\alpha$ with time, $t$, for different $C_{ratio}$ values. The color code is same the for both the insets.}   
    \label{fig:5}
\end{figure*}

{\em Cluster size distributions of bacteria} : 
The cluster size distributions (CSD), $P(n)$ $vs.$ $n$, of bacteria for different values of $C_{ratio}$ in FIG.\ref{fig:5}. For $C_{ratio}$ values in the $f > 0$ region, the distribution is short tailed and exhibits algebraic decay for $n > 2$ : $P(n) \sim n^{-\alpha}$, where $\alpha > 1$ as shown in the bottom left inset of FIG.\ref{fig:5}(b). The main frame in Fig.\ref{fig:5}(b) showcases the $P(n)$ $vs.$ $n$ plot at different times, wherein it is evident that due to efficient engulfment of bacteria the decay of $P(n)$ sharpens over time. The top-right inset in FIG.\ref{fig:5}(b) depicts the time variation variation of $\alpha$. Moreover, the short-tailed form of the distribution indicates a low probability of encountering larger clusters of bacteria within the system, while the algebraic decay signals a scale-free structure. The scale free nature of the system suggests that the relative probability of the cluster sizes remain statistically invariant irrespective of the actual size of the system i.e. $\frac{P(n_2)}{P(n_1)} = \frac{P(n_4)}{P(n_3)} = k = \text{constant} >0$ where, $n_2 = k n_1$, and $n_4 = k n_3$. Thus, $P(n_2) = k^{-\alpha} P(n_1)$. For $\alpha >1$, this implies that successively larger cluster sizes ($n > 2$)  are equally less probable.\\
For $C_{ratio}$ values in $f < 0$ regime,  the distribution shows a plateau region for small n, $n \approx 3 - 7$, followed by an exponential decay with a fat tail and small secondary peak for large $n$ ($\approx 40 - 50$). The exponential decay of $P(n)$ for the intermediate range of $n$ is visible in the bottom-left inset in FIG.\ref{fig:5}(a) wherein the $P(n)$ $vs.$ $n$ plot is shown on a log-y scale. The main frame in FIG.\ref{fig:5}(a) displays the plot of $P(n)$ $vs.$ $n$ at different times, wherein it is evident that due to reproduction of bacteria and inefficient engulfment, the probability of encountering a large bacteria cluster increases with time. The plateau implies the enhanced probability of formation of clusters of these sizes. Further the fat tail of the distribution and the secondary peak at large $n$ suggest a finite probability of observing larger clusters of 40–50 bacteria. However, very large clusters are exponentially less likely to find compared to the smaller ones.\\ 

\begin{figure}[hbt]
    \centering
    \includegraphics[width=0.995\linewidth]{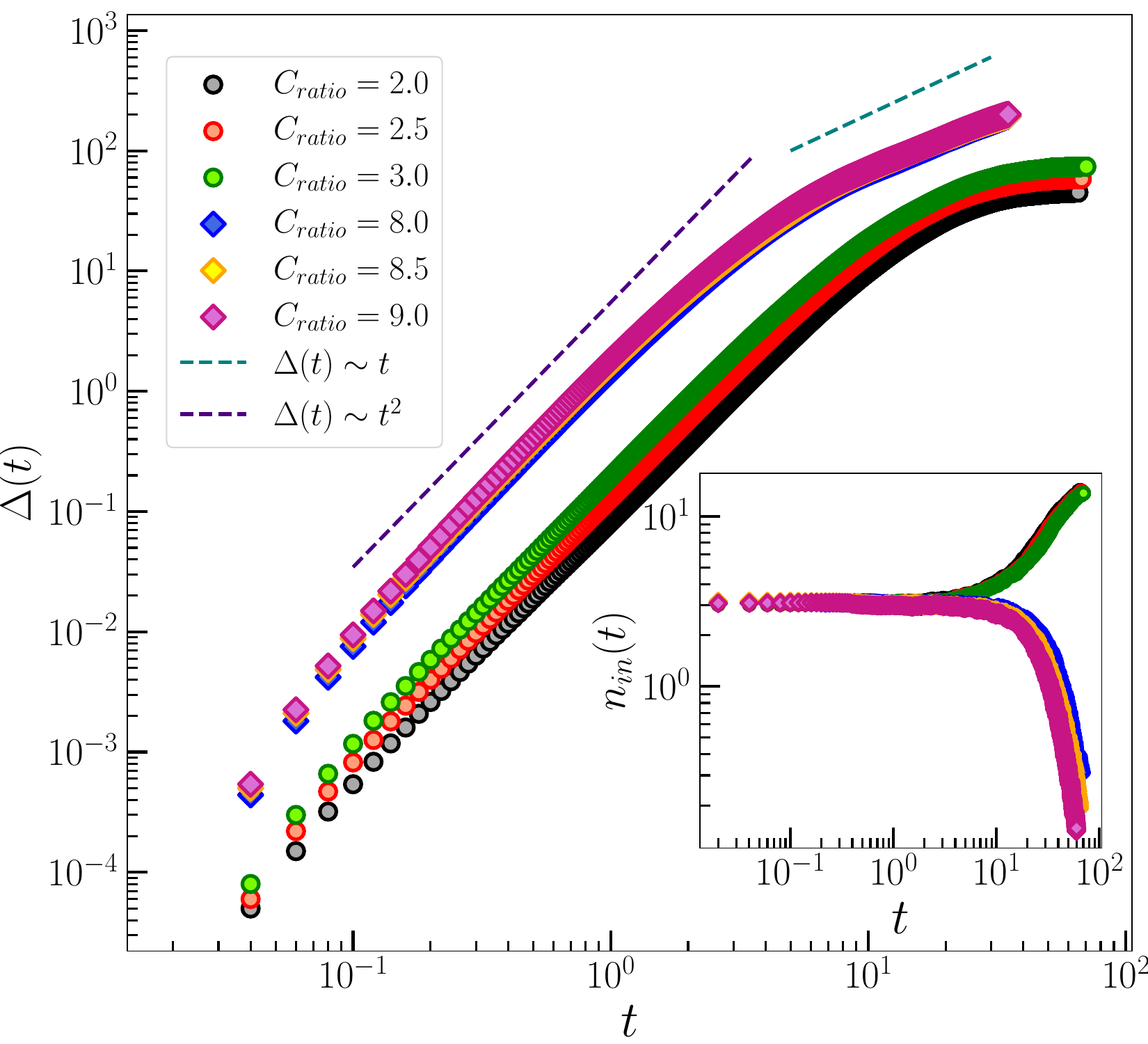}
    \caption{(color online) The figure displays the behavior of mean square displacement of phagocytes, $\Delta(t)$ $vs.$ $t$ for differnt values of $C_{ratio}$ in $f > 0$ ($C_{ratio} =$ $3.5$, $4.0$, $\&$ $4.5$) and $f < 0$ regions ($C_{ratio} =$ $8.0$, $8.5$, $\&$ $9.0$) in the main frame. The inset shows the plot of $n_{in}(t)$ $vs.$ $t$ for different values of $C_{ratio}$. The plots in the main frame and the inset are shown in log-log scale. The legends for different values of $C_{ratio}$ are consistent for main frame and inset. }
    \label{fig:6}
\end{figure}

{\em Dynamic of Phagocytes across the Phase Boundary} : To characterize the dynamics of phagocytes, first we calculate their mean square displacement (MSD) defined as 
\begin{equation*}
\Delta (t) = \bigg< \dfrac{1}{N_p} \sum_{i=1}^{N_{p}} |\boldsymbol{r}_{i,p}(t_0 + t)-\boldsymbol{r}_{i,p}(t_0)|^2 \bigg> 	  
\end{equation*}
where, $<....>$ implies an average over independent realizations. The variation of MSD with time can be expressed as a powerlaw, $\Delta(t) \sim t^{\beta}$. The exponent $\beta$ carries the information regarding the nature of motion exhibited by particles : $\beta = 1$ $\&$ $2$ corresponds to diffusive and ballistic motion, respectively, whereas $\beta < 1$ and $1<\beta<2$ implies sub diffusive and super diffusive motion, respectively.\\
The plots of MSD for different values of $C_{ratio}$ is shown in FIG.\ref{fig:6}. The behavior of $\Delta(t)$ $vs.$ $t$ plot for all the values of $C_{ratio}$ is qualitatively similar for all the $C_{ratio}$ values at very early time wherein phagocytes exhibit ballistic dynamic i.e. $\Delta(t) \sim t^2$. However, the late time behavior of $\Delta(t)$ is substantially different for $C_{ratio}$ values in $f > 0$ and $f < 0$ regime of the phase diagram. In $f > 0$ regime, the graph crosses over to a sub diffusive regime with $\beta \approx 0.90$. Whereas for $C_{ratio}$ values in the $f < 0$ regime, $\Delta(t)$ exhibits a smooth cross over from early time ballistic regime to a late time saturation regime.\\
To explain the underlying mechanism leading to the late time characteristics of $\Delta(t)$ in $f > 0$ and $f < 0$ regime, we calculate the average number of bacteria inside the interaction radius of a phagocyte, denoted by $n_{in}(t)$. The plot of $n_{in}(t)$ $vs.$ $t$ is shown in the inset of FIG.\ref{fig:6}(a). The plot clearly shows that during early time the average number of bacteria inside the interaction radius of phagocytes are very small and it stays nearly constant. In this regime the parameters in the $f > 0$ and $f > 0$ regime are equivalent. This leads to the early time balletic behavior of $\Delta(t)$ in both the regimes. Beyond this regime, the value $n(t)$ increases rapidly for $C_{ratio}$ values in $f < 0$ regime. This causes to trapping of phagocytes and hence, $\Delta(t)$ saturates. On the contrary, for $C_{ratio}$ values in $f > 0$ regime, $n(t)$ decreases  with time. This leads to the ballistic to the sub-diffusive cross over in $\Delta(t)$. In the sub-diffusive regime, the motion of phagocytes varies based on the number of bacteria within their interaction radius. Phagocytes with a higher number of bacteria within their interaction radius exhibit motion faster than diffusion, while those with no bacteria experience slower-than-diffusive motion due to overdamped dynamics.  The combined effect of these two leads to sub-diffusive dynamics of when averaged over all the phagocytes in the system. \\

{\section{Discussion}\label{sec:dis}}
To summarize, this study introduces a minimal model to explore phagocytosis in a multi-agent system composed of bacteria and phagocytes. Bacteria are depicted as self-propelled disks, while phagocytes are modeled as passive disks much larger in size compared to bacteria. Interactions within each species follow volume exclusion rules, while inter species interactions are governed by two main forces: a repulsive force that drives bacteria away from phagocytes, and an attractive force that draws phagocytes toward bacteria. Additionally, bacteria reproduce at a fixed rate, creating a dynamic balance between bacterial growth and phagocytic engulfment. This balance, influenced by the competition between the opposing forces in inter species interactions, ultimately determines the fate of the system. Consequently, the relative range and strength of attraction and repulsion, denoted by $C_{ratio}$ and $A_{ratio}$, respectively, serve as the control parameters of the model.\\
To assess phagocytosis efficiency, we define $f$ as the fraction of bacteria remaining over time relative to the initial count. Depending on control parameters, the system stabilizes into either an $f > 0$ state, where population of of bacteria decrease over time, or an $f < 0$ state, where population of bacteria bacteria grow. A phase diagram based on the probability of positive phagocytosis, $P(f>0)$, shows two distinct regimes with a gradual transition through a bi-stable region, where system behavior varies across ensembles. The analysis of cluster size distribution (CSD) of bacteria and mean square displacement (MSD) of phagocytes reveals the following : in the $f > 0$ regime bacteria form large clusters that trap phagocytes, inhibiting phagocytosis. Conversely, in the $f > 0$ regime, strong phagocyte-bacteria attraction enables effective engulfment, keeping bacterial numbers low, and preventing large bacteria clusters from forming\\
In closing, we recognize several limitations of our study. Firstly, we modeled both the bacteria and phagocytes as circular disks, whereas in vitro experiments show that target shape can significantly affect phagocytosis efficiency \cite{paul2013phagocytosis}. Secondly, our model does not account for the deterioration of phagocyte health during interactions with bacteria, which has been observed experimentally. Examining the impact of these aspects on our results would be an interesting direction for future research.\\

\section{Acknowledgement}
The authors thank Prof. Manoranjan Kumar for useful discussions. P.S.M., P.K.M. and S.M., thanks PARAM Shivay for computational facility under the National Supercomputing Mission, Government of India at the Indian Institute of Technology, Varanasi and the computational facility at I.I.T. (BHU) Varanasi. P.S.M. and P.K.M.  thank UGC for research fellowship. S.M. thanks DST, SERB (INDIA), Project No.: CRG/2021/006945, MTR/2021/000438  for financial support.\\

\bibliography{references}

\begin{thebibliography}{27}%
\makeatletter
\providecommand \@ifxundefined [1]{%
 \@ifx{#1\undefined}
}%
\providecommand \@ifnum [1]{%
 \ifnum #1\expandafter \@firstoftwo
 \else \expandafter \@secondoftwo
 \fi
}%
\providecommand \@ifx [1]{%
 \ifx #1\expandafter \@firstoftwo
 \else \expandafter \@secondoftwo
 \fi
}%
\providecommand \natexlab [1]{#1}%
\providecommand \enquote  [1]{``#1''}%
\providecommand \bibnamefont  [1]{#1}%
\providecommand \bibfnamefont [1]{#1}%
\providecommand \citenamefont [1]{#1}%
\providecommand \href@noop [0]{\@secondoftwo}%
\providecommand \href [0]{\begingroup \@sanitize@url \@href}%
\providecommand \@href[1]{\@@startlink{#1}\@@href}%
\providecommand \@@href[1]{\endgroup#1\@@endlink}%
\providecommand \@sanitize@url [0]{\catcode `\\12\catcode `\$12\catcode `\&12\catcode `\#12\catcode `\^12\catcode `\_12\catcode `\%12\relax}%
\providecommand \@@startlink[1]{}%
\providecommand \@@endlink[0]{}%
\providecommand \url  [0]{\begingroup\@sanitize@url \@url }%
\providecommand \@url [1]{\endgroup\@href {#1}{\urlprefix }}%
\providecommand \urlprefix  [0]{URL }%
\providecommand \Eprint [0]{\href }%
\providecommand \doibase [0]{https://doi.org/}%
\providecommand \selectlanguage [0]{\@gobble}%
\providecommand \bibinfo  [0]{\@secondoftwo}%
\providecommand \bibfield  [0]{\@secondoftwo}%
\providecommand \translation [1]{[#1]}%
\providecommand \BibitemOpen [0]{}%
\providecommand \bibitemStop [0]{}%
\providecommand \bibitemNoStop [0]{.\EOS\space}%
\providecommand \EOS [0]{\spacefactor3000\relax}%
\providecommand \BibitemShut  [1]{\csname bibitem#1\endcsname}%
\let\auto@bib@innerbib\@empty
\bibitem [{\citenamefont {Vicsek}\ \emph {et~al.}(1995)\citenamefont {Vicsek}, \citenamefont {Czir{\'o}k}, \citenamefont {Ben-Jacob}, \citenamefont {Cohen},\ and\ \citenamefont {Shochet}}]{vicsek1995novel}%
  \BibitemOpen
  \bibfield  {author} {\bibinfo {author} {\bibfnamefont {T.}~\bibnamefont {Vicsek}}, \bibinfo {author} {\bibfnamefont {A.}~\bibnamefont {Czir{\'o}k}}, \bibinfo {author} {\bibfnamefont {E.}~\bibnamefont {Ben-Jacob}}, \bibinfo {author} {\bibfnamefont {I.}~\bibnamefont {Cohen}},\ and\ \bibinfo {author} {\bibfnamefont {O.}~\bibnamefont {Shochet}},\ }\bibfield  {title} {\bibinfo {title} {Novel type of phase transition in a system of self-driven particles},\ }\href@noop {} {\bibfield  {journal} {\bibinfo  {journal} {Physical review letters}\ }\textbf {\bibinfo {volume} {75}},\ \bibinfo {pages} {1226} (\bibinfo {year} {1995})}\BibitemShut {NoStop}%
\bibitem [{\citenamefont {Bechinger}\ \emph {et~al.}(2016)\citenamefont {Bechinger}, \citenamefont {Di~Leonardo}, \citenamefont {L{\"o}wen}, \citenamefont {Reichhardt}, \citenamefont {Volpe},\ and\ \citenamefont {Volpe}}]{bechinger2016active}%
  \BibitemOpen
  \bibfield  {author} {\bibinfo {author} {\bibfnamefont {C.}~\bibnamefont {Bechinger}}, \bibinfo {author} {\bibfnamefont {R.}~\bibnamefont {Di~Leonardo}}, \bibinfo {author} {\bibfnamefont {H.}~\bibnamefont {L{\"o}wen}}, \bibinfo {author} {\bibfnamefont {C.}~\bibnamefont {Reichhardt}}, \bibinfo {author} {\bibfnamefont {G.}~\bibnamefont {Volpe}},\ and\ \bibinfo {author} {\bibfnamefont {G.}~\bibnamefont {Volpe}},\ }\bibfield  {title} {\bibinfo {title} {Active particles in complex and crowded environments},\ }\href@noop {} {\bibfield  {journal} {\bibinfo  {journal} {Reviews of modern physics}\ }\textbf {\bibinfo {volume} {88}},\ \bibinfo {pages} {045006} (\bibinfo {year} {2016})}\BibitemShut {NoStop}%
\bibitem [{\citenamefont {B{\"a}r}\ \emph {et~al.}(2020)\citenamefont {B{\"a}r}, \citenamefont {Gro{\ss}mann}, \citenamefont {Heidenreich},\ and\ \citenamefont {Peruani}}]{bar2020self}%
  \BibitemOpen
  \bibfield  {author} {\bibinfo {author} {\bibfnamefont {M.}~\bibnamefont {B{\"a}r}}, \bibinfo {author} {\bibfnamefont {R.}~\bibnamefont {Gro{\ss}mann}}, \bibinfo {author} {\bibfnamefont {S.}~\bibnamefont {Heidenreich}},\ and\ \bibinfo {author} {\bibfnamefont {F.}~\bibnamefont {Peruani}},\ }\bibfield  {title} {\bibinfo {title} {Self-propelled rods: Insights and perspectives for active matter},\ }\href@noop {} {\bibfield  {journal} {\bibinfo  {journal} {Annual Review of Condensed Matter Physics}\ }\textbf {\bibinfo {volume} {11}},\ \bibinfo {pages} {441} (\bibinfo {year} {2020})}\BibitemShut {NoStop}%
\bibitem [{\citenamefont {Vicsek}\ and\ \citenamefont {Zafeiris}(2012)}]{vicsek2012collective}%
  \BibitemOpen
  \bibfield  {author} {\bibinfo {author} {\bibfnamefont {T.}~\bibnamefont {Vicsek}}\ and\ \bibinfo {author} {\bibfnamefont {A.}~\bibnamefont {Zafeiris}},\ }\bibfield  {title} {\bibinfo {title} {Collective motion},\ }\href@noop {} {\bibfield  {journal} {\bibinfo  {journal} {Physics reports}\ }\textbf {\bibinfo {volume} {517}},\ \bibinfo {pages} {71} (\bibinfo {year} {2012})}\BibitemShut {NoStop}%
\bibitem [{\citenamefont {Toner}\ \emph {et~al.}(2005)\citenamefont {Toner}, \citenamefont {Tu},\ and\ \citenamefont {Ramaswamy}}]{toner2005hydrodynamics}%
  \BibitemOpen
  \bibfield  {author} {\bibinfo {author} {\bibfnamefont {J.}~\bibnamefont {Toner}}, \bibinfo {author} {\bibfnamefont {Y.}~\bibnamefont {Tu}},\ and\ \bibinfo {author} {\bibfnamefont {S.}~\bibnamefont {Ramaswamy}},\ }\bibfield  {title} {\bibinfo {title} {Hydrodynamics and phases of flocks},\ }\href@noop {} {\bibfield  {journal} {\bibinfo  {journal} {Annals of Physics}\ }\textbf {\bibinfo {volume} {318}},\ \bibinfo {pages} {170} (\bibinfo {year} {2005})}\BibitemShut {NoStop}%
\bibitem [{\citenamefont {Marchetti}\ \emph {et~al.}(2013)\citenamefont {Marchetti}, \citenamefont {Joanny}, \citenamefont {Ramaswamy}, \citenamefont {Liverpool}, \citenamefont {Prost}, \citenamefont {Rao},\ and\ \citenamefont {Simha}}]{marchetti2013hydrodynamics}%
  \BibitemOpen
  \bibfield  {author} {\bibinfo {author} {\bibfnamefont {M.~C.}\ \bibnamefont {Marchetti}}, \bibinfo {author} {\bibfnamefont {J.-F.}\ \bibnamefont {Joanny}}, \bibinfo {author} {\bibfnamefont {S.}~\bibnamefont {Ramaswamy}}, \bibinfo {author} {\bibfnamefont {T.~B.}\ \bibnamefont {Liverpool}}, \bibinfo {author} {\bibfnamefont {J.}~\bibnamefont {Prost}}, \bibinfo {author} {\bibfnamefont {M.}~\bibnamefont {Rao}},\ and\ \bibinfo {author} {\bibfnamefont {R.~A.}\ \bibnamefont {Simha}},\ }\bibfield  {title} {\bibinfo {title} {Hydrodynamics of soft active matter},\ }\href@noop {} {\bibfield  {journal} {\bibinfo  {journal} {Reviews of modern physics}\ }\textbf {\bibinfo {volume} {85}},\ \bibinfo {pages} {1143} (\bibinfo {year} {2013})}\BibitemShut {NoStop}%
\bibitem [{\citenamefont {Ramaswamy}(2017)}]{ramaswamy2017active}%
  \BibitemOpen
  \bibfield  {author} {\bibinfo {author} {\bibfnamefont {S.}~\bibnamefont {Ramaswamy}},\ }\bibfield  {title} {\bibinfo {title} {Active matter},\ }\href@noop {} {\bibfield  {journal} {\bibinfo  {journal} {Journal of Statistical Mechanics: Theory and Experiment}\ }\textbf {\bibinfo {volume} {2017}},\ \bibinfo {pages} {054002} (\bibinfo {year} {2017})}\BibitemShut {NoStop}%
\bibitem [{\citenamefont {Das}\ \emph {et~al.}(2020)\citenamefont {Das}, \citenamefont {Schmidt},\ and\ \citenamefont {Murrell}}]{das2020introduction}%
  \BibitemOpen
  \bibfield  {author} {\bibinfo {author} {\bibfnamefont {M.}~\bibnamefont {Das}}, \bibinfo {author} {\bibfnamefont {C.~F.}\ \bibnamefont {Schmidt}},\ and\ \bibinfo {author} {\bibfnamefont {M.}~\bibnamefont {Murrell}},\ }\bibfield  {title} {\bibinfo {title} {Introduction to active matter},\ }\href@noop {} {\bibfield  {journal} {\bibinfo  {journal} {Soft Matter}\ }\textbf {\bibinfo {volume} {16}},\ \bibinfo {pages} {7185} (\bibinfo {year} {2020})}\BibitemShut {NoStop}%
\bibitem [{\citenamefont {Toner}(2012)}]{toner2012birth}%
  \BibitemOpen
  \bibfield  {author} {\bibinfo {author} {\bibfnamefont {J.}~\bibnamefont {Toner}},\ }\bibfield  {title} {\bibinfo {title} {Birth, death, and flight: A theory of malthusian flocks},\ }\href@noop {} {\bibfield  {journal} {\bibinfo  {journal} {Physical review letters}\ }\textbf {\bibinfo {volume} {108}},\ \bibinfo {pages} {088102} (\bibinfo {year} {2012})}\BibitemShut {NoStop}%
\bibitem [{\citenamefont {Mishra}\ and\ \citenamefont {Mishra}(2022)}]{mishra2022active}%
  \BibitemOpen
  \bibfield  {author} {\bibinfo {author} {\bibfnamefont {P.~K.}\ \bibnamefont {Mishra}}\ and\ \bibinfo {author} {\bibfnamefont {S.}~\bibnamefont {Mishra}},\ }\bibfield  {title} {\bibinfo {title} {Active polar flock with birth and death},\ }\href@noop {} {\bibfield  {journal} {\bibinfo  {journal} {Physics of Fluids}\ }\textbf {\bibinfo {volume} {34}} (\bibinfo {year} {2022})}\BibitemShut {NoStop}%
\bibitem [{\citenamefont {Jena}\ and\ \citenamefont {Mishra}(2023)}]{jena2023ordering}%
  \BibitemOpen
  \bibfield  {author} {\bibinfo {author} {\bibfnamefont {A.~P.}\ \bibnamefont {Jena}}\ and\ \bibinfo {author} {\bibfnamefont {B.~S.}\ \bibnamefont {Mishra}},\ }\bibfield  {title} {\bibinfo {title} {Ordering kinetics and steady state of malthusian flock},\ }\href@noop {} {\bibfield  {journal} {\bibinfo  {journal} {Physics of Fluids}\ }\textbf {\bibinfo {volume} {35}} (\bibinfo {year} {2023})}\BibitemShut {NoStop}%
\bibitem [{\citenamefont {Jena}\ and\ \citenamefont {Mishra}(2024)}]{jena2024spatio}%
  \BibitemOpen
  \bibfield  {author} {\bibinfo {author} {\bibfnamefont {P.}~\bibnamefont {Jena}}\ and\ \bibinfo {author} {\bibfnamefont {S.}~\bibnamefont {Mishra}},\ }\bibfield  {title} {\bibinfo {title} {Spatio-temporal patterns in growing bacterial suspensions: Impact of growth dynamics},\ }\href@noop {} {\bibfield  {journal} {\bibinfo  {journal} {arXiv preprint arXiv:2408.00403}\ } (\bibinfo {year} {2024})}\BibitemShut {NoStop}%
\bibitem [{\citenamefont {Mondal}\ \emph {et~al.}(2024)\citenamefont {Mondal}, \citenamefont {Mishra}, \citenamefont {Vicsek},\ and\ \citenamefont {Mishra}}]{mondal2024dynamical}%
  \BibitemOpen
  \bibfield  {author} {\bibinfo {author} {\bibfnamefont {P.~S.}\ \bibnamefont {Mondal}}, \bibinfo {author} {\bibfnamefont {P.~K.}\ \bibnamefont {Mishra}}, \bibinfo {author} {\bibfnamefont {T.}~\bibnamefont {Vicsek}},\ and\ \bibinfo {author} {\bibfnamefont {S.}~\bibnamefont {Mishra}},\ }\bibfield  {title} {\bibinfo {title} {Dynamical swirl structures powered by microswimmers in active nematics},\ }\href@noop {} {\bibfield  {journal} {\bibinfo  {journal} {arXiv preprint arXiv:2407.05861}\ } (\bibinfo {year} {2024})}\BibitemShut {NoStop}%
\bibitem [{\citenamefont {Mishra}\ \emph {et~al.}(2024)\citenamefont {Mishra}, \citenamefont {Puitandy},\ and\ \citenamefont {Mishra}}]{Mishra_2024}%
  \BibitemOpen
  \bibfield  {author} {\bibinfo {author} {\bibfnamefont {P.~K.}\ \bibnamefont {Mishra}}, \bibinfo {author} {\bibfnamefont {A.}~\bibnamefont {Puitandy}},\ and\ \bibinfo {author} {\bibfnamefont {S.}~\bibnamefont {Mishra}},\ }\bibfield  {title} {\bibinfo {title} {Directional cues affect the collective behaviour of self-propelled particles in one dimension},\ }\href {https://doi.org/10.1209/0295-5075/ad749c} {\bibfield  {journal} {\bibinfo  {journal} {Europhysics Letters}\ }\textbf {\bibinfo {volume} {147}},\ \bibinfo {pages} {67001} (\bibinfo {year} {2024})}\BibitemShut {NoStop}%
\bibitem [{\citenamefont {Sampat}\ and\ \citenamefont {Mishra}(2021)}]{sampat2021polar}%
  \BibitemOpen
  \bibfield  {author} {\bibinfo {author} {\bibfnamefont {P.~B.}\ \bibnamefont {Sampat}}\ and\ \bibinfo {author} {\bibfnamefont {S.}~\bibnamefont {Mishra}},\ }\bibfield  {title} {\bibinfo {title} {Polar swimmers induce several phases in active nematics},\ }\href@noop {} {\bibfield  {journal} {\bibinfo  {journal} {Physical Review E}\ }\textbf {\bibinfo {volume} {104}},\ \bibinfo {pages} {024130} (\bibinfo {year} {2021})}\BibitemShut {NoStop}%
\bibitem [{\citenamefont {McCandlish}\ \emph {et~al.}(2012)\citenamefont {McCandlish}, \citenamefont {Baskaran},\ and\ \citenamefont {Hagan}}]{mccandlish2012spontaneous}%
  \BibitemOpen
  \bibfield  {author} {\bibinfo {author} {\bibfnamefont {S.~R.}\ \bibnamefont {McCandlish}}, \bibinfo {author} {\bibfnamefont {A.}~\bibnamefont {Baskaran}},\ and\ \bibinfo {author} {\bibfnamefont {M.~F.}\ \bibnamefont {Hagan}},\ }\bibfield  {title} {\bibinfo {title} {Spontaneous segregation of self-propelled particles with different motilities},\ }\href@noop {} {\bibfield  {journal} {\bibinfo  {journal} {Soft Matter}\ }\textbf {\bibinfo {volume} {8}},\ \bibinfo {pages} {2527} (\bibinfo {year} {2012})}\BibitemShut {NoStop}%
\bibitem [{\citenamefont {Singh}\ \emph {et~al.}(2021)\citenamefont {Singh}, \citenamefont {Kumar},\ and\ \citenamefont {Mishra}}]{singh2021bond}%
  \BibitemOpen
  \bibfield  {author} {\bibinfo {author} {\bibfnamefont {J.~P.}\ \bibnamefont {Singh}}, \bibinfo {author} {\bibfnamefont {S.}~\bibnamefont {Kumar}},\ and\ \bibinfo {author} {\bibfnamefont {S.}~\bibnamefont {Mishra}},\ }\bibfield  {title} {\bibinfo {title} {Bond disorder enhances the information transfer in the polar flock},\ }\href@noop {} {\bibfield  {journal} {\bibinfo  {journal} {Journal of Statistical Mechanics: Theory and Experiment}\ }\textbf {\bibinfo {volume} {2021}},\ \bibinfo {pages} {083217} (\bibinfo {year} {2021})}\BibitemShut {NoStop}%
\bibitem [{\citenamefont {Chepizhko}\ \emph {et~al.}(2013)\citenamefont {Chepizhko}, \citenamefont {Altmann},\ and\ \citenamefont {Peruani}}]{chepizhko2013optimal}%
  \BibitemOpen
  \bibfield  {author} {\bibinfo {author} {\bibfnamefont {O.}~\bibnamefont {Chepizhko}}, \bibinfo {author} {\bibfnamefont {E.~G.}\ \bibnamefont {Altmann}},\ and\ \bibinfo {author} {\bibfnamefont {F.}~\bibnamefont {Peruani}},\ }\bibfield  {title} {\bibinfo {title} {Optimal noise maximizes collective motion in heterogeneous media},\ }\href@noop {} {\bibfield  {journal} {\bibinfo  {journal} {Physical review letters}\ }\textbf {\bibinfo {volume} {110}},\ \bibinfo {pages} {238101} (\bibinfo {year} {2013})}\BibitemShut {NoStop}%
\bibitem [{\citenamefont {Das}\ \emph {et~al.}(2018)\citenamefont {Das}, \citenamefont {Kumar},\ and\ \citenamefont {Mishra}}]{das2018polar}%
  \BibitemOpen
  \bibfield  {author} {\bibinfo {author} {\bibfnamefont {R.}~\bibnamefont {Das}}, \bibinfo {author} {\bibfnamefont {M.}~\bibnamefont {Kumar}},\ and\ \bibinfo {author} {\bibfnamefont {S.}~\bibnamefont {Mishra}},\ }\bibfield  {title} {\bibinfo {title} {Polar flock in the presence of random quenched rotators},\ }\href@noop {} {\bibfield  {journal} {\bibinfo  {journal} {Physical Review E}\ }\textbf {\bibinfo {volume} {98}},\ \bibinfo {pages} {060602} (\bibinfo {year} {2018})}\BibitemShut {NoStop}%
\bibitem [{\citenamefont {Dolai}\ \emph {et~al.}(2018)\citenamefont {Dolai}, \citenamefont {Simha},\ and\ \citenamefont {Mishra}}]{dolai2018phase}%
  \BibitemOpen
  \bibfield  {author} {\bibinfo {author} {\bibfnamefont {P.}~\bibnamefont {Dolai}}, \bibinfo {author} {\bibfnamefont {A.}~\bibnamefont {Simha}},\ and\ \bibinfo {author} {\bibfnamefont {S.}~\bibnamefont {Mishra}},\ }\bibfield  {title} {\bibinfo {title} {Phase separation in binary mixtures of active and passive particles},\ }\href@noop {} {\bibfield  {journal} {\bibinfo  {journal} {Soft Matter}\ }\textbf {\bibinfo {volume} {14}},\ \bibinfo {pages} {6137} (\bibinfo {year} {2018})}\BibitemShut {NoStop}%
\bibitem [{\citenamefont {Uribe-Querol}\ and\ \citenamefont {Rosales}(2020)}]{uribe2020phagocytosis}%
  \BibitemOpen
  \bibfield  {author} {\bibinfo {author} {\bibfnamefont {E.}~\bibnamefont {Uribe-Querol}}\ and\ \bibinfo {author} {\bibfnamefont {C.}~\bibnamefont {Rosales}},\ }\bibfield  {title} {\bibinfo {title} {Phagocytosis: our current understanding of a universal biological process},\ }\href@noop {} {\bibfield  {journal} {\bibinfo  {journal} {Frontiers in immunology}\ }\textbf {\bibinfo {volume} {11}},\ \bibinfo {pages} {1066} (\bibinfo {year} {2020})}\BibitemShut {NoStop}%
\bibitem [{\citenamefont {Richards}\ and\ \citenamefont {Endres}(2014)}]{richards2014mechanism}%
  \BibitemOpen
  \bibfield  {author} {\bibinfo {author} {\bibfnamefont {D.~M.}\ \bibnamefont {Richards}}\ and\ \bibinfo {author} {\bibfnamefont {R.~G.}\ \bibnamefont {Endres}},\ }\bibfield  {title} {\bibinfo {title} {The mechanism of phagocytosis: two stages of engulfment},\ }\href@noop {} {\bibfield  {journal} {\bibinfo  {journal} {Biophysical journal}\ }\textbf {\bibinfo {volume} {107}},\ \bibinfo {pages} {1542} (\bibinfo {year} {2014})}\BibitemShut {NoStop}%
\bibitem [{\citenamefont {Herant}\ \emph {et~al.}(2006)\citenamefont {Herant}, \citenamefont {Heinrich},\ and\ \citenamefont {Dembo}}]{herant2006mechanics}%
  \BibitemOpen
  \bibfield  {author} {\bibinfo {author} {\bibfnamefont {M.}~\bibnamefont {Herant}}, \bibinfo {author} {\bibfnamefont {V.}~\bibnamefont {Heinrich}},\ and\ \bibinfo {author} {\bibfnamefont {M.}~\bibnamefont {Dembo}},\ }\bibfield  {title} {\bibinfo {title} {Mechanics of neutrophil phagocytosis: experiments and quantitative models},\ }\href@noop {} {\bibfield  {journal} {\bibinfo  {journal} {Journal of cell science}\ }\textbf {\bibinfo {volume} {119}},\ \bibinfo {pages} {1903} (\bibinfo {year} {2006})}\BibitemShut {NoStop}%
\bibitem [{\citenamefont {Sadhu}\ \emph {et~al.}(2023)\citenamefont {Sadhu}, \citenamefont {Barger}, \citenamefont {Peni{\v{c}}}, \citenamefont {Igli{\v{c}}}, \citenamefont {Krendel}, \citenamefont {Gauthier},\ and\ \citenamefont {Gov}}]{sadhu2023theoretical}%
  \BibitemOpen
  \bibfield  {author} {\bibinfo {author} {\bibfnamefont {R.~K.}\ \bibnamefont {Sadhu}}, \bibinfo {author} {\bibfnamefont {S.~R.}\ \bibnamefont {Barger}}, \bibinfo {author} {\bibfnamefont {S.}~\bibnamefont {Peni{\v{c}}}}, \bibinfo {author} {\bibfnamefont {A.}~\bibnamefont {Igli{\v{c}}}}, \bibinfo {author} {\bibfnamefont {M.}~\bibnamefont {Krendel}}, \bibinfo {author} {\bibfnamefont {N.~C.}\ \bibnamefont {Gauthier}},\ and\ \bibinfo {author} {\bibfnamefont {N.~S.}\ \bibnamefont {Gov}},\ }\bibfield  {title} {\bibinfo {title} {A theoretical model of efficient phagocytosis driven by curved membrane proteins and active cytoskeleton forces},\ }\href@noop {} {\bibfield  {journal} {\bibinfo  {journal} {Soft matter}\ }\textbf {\bibinfo {volume} {19}},\ \bibinfo {pages} {31} (\bibinfo {year} {2023})}\BibitemShut {NoStop}%
\bibitem [{\citenamefont {Hamed}\ and\ \citenamefont {Nepomnyashchy}(2021)}]{hamed2021three}%
  \BibitemOpen
  \bibfield  {author} {\bibinfo {author} {\bibfnamefont {M.~A.}\ \bibnamefont {Hamed}}\ and\ \bibinfo {author} {\bibfnamefont {A.~A.}\ \bibnamefont {Nepomnyashchy}},\ }\bibfield  {title} {\bibinfo {title} {Three-dimensional phase field model for actin-based cell membrane dynamics},\ }\href@noop {} {\bibfield  {journal} {\bibinfo  {journal} {Mathematical Modelling of Natural Phenomena}\ }\textbf {\bibinfo {volume} {16}},\ \bibinfo {pages} {56} (\bibinfo {year} {2021})}\BibitemShut {NoStop}%
\bibitem [{\citenamefont {Winkler}\ \emph {et~al.}(2024)\citenamefont {Winkler}, \citenamefont {Hamed}, \citenamefont {Nepomnyashchy},\ and\ \citenamefont {Ziebert}}]{winkler2024physical}%
  \BibitemOpen
  \bibfield  {author} {\bibinfo {author} {\bibfnamefont {B.}~\bibnamefont {Winkler}}, \bibinfo {author} {\bibfnamefont {M.~A.}\ \bibnamefont {Hamed}}, \bibinfo {author} {\bibfnamefont {A.~A.}\ \bibnamefont {Nepomnyashchy}},\ and\ \bibinfo {author} {\bibfnamefont {F.}~\bibnamefont {Ziebert}},\ }\bibfield  {title} {\bibinfo {title} {Physical phase field model for phagocytosis},\ }\href@noop {} {\bibfield  {journal} {\bibinfo  {journal} {New Journal of Physics}\ }\textbf {\bibinfo {volume} {26}},\ \bibinfo {pages} {013029} (\bibinfo {year} {2024})}\BibitemShut {NoStop}%
\bibitem [{\citenamefont {Paul}\ \emph {et~al.}(2013)\citenamefont {Paul}, \citenamefont {Achouri}, \citenamefont {Yoon}, \citenamefont {Herre}, \citenamefont {Bryant},\ and\ \citenamefont {Cicuta}}]{paul2013phagocytosis}%
  \BibitemOpen
  \bibfield  {author} {\bibinfo {author} {\bibfnamefont {D.}~\bibnamefont {Paul}}, \bibinfo {author} {\bibfnamefont {S.}~\bibnamefont {Achouri}}, \bibinfo {author} {\bibfnamefont {Y.-Z.}\ \bibnamefont {Yoon}}, \bibinfo {author} {\bibfnamefont {J.}~\bibnamefont {Herre}}, \bibinfo {author} {\bibfnamefont {C.~E.}\ \bibnamefont {Bryant}},\ and\ \bibinfo {author} {\bibfnamefont {P.}~\bibnamefont {Cicuta}},\ }\bibfield  {title} {\bibinfo {title} {Phagocytosis dynamics depends on target shape},\ }\href@noop {} {\bibfield  {journal} {\bibinfo  {journal} {Biophysical journal}\ }\textbf {\bibinfo {volume} {105}},\ \bibinfo {pages} {1143} (\bibinfo {year} {2013})}\BibitemShut {NoStop}%
\end{thebibliography}%

\end{document}